# Reemerging superconductivity at 48 K in iron chalcogenides

Liling Sun[1,8], Xiao-Jia Chen[2,3,8], Jing Guo[1], Peiwen Gao[1], Qing-Zhen Huang[4], Hangdong Wang[5], Minghu Fang[5], Xiaolong Chen[1], Genfu Chen[1], Qi Wu[1], Chao Zhang[1], Dachun Gu[1], Xiaoli Dong[1], Lin Wang[6], Ke Yang[7], Aiguo Li[7], Xi Dai[1], Ho-kwang Mao[2] & Zhongxian Zhao[1]

[1]*Institute of Physics and Beijing National Laboratory for Condensed Matter Physics, Chinese Academy of Sciences, Beijing 100190, China*

[2]*Geophysical Laboratory, Carnegie Institution of Washington, Washington, D.C. 20015, USA*

[3]*Department of Physics, South China University of Technology, Guangzhou 510640, China*

[4]*NIST Center for Neutron Research, National Institute of Standards and Technology, Gaithersburg, MD 20899, USA*

[5]*Department of Physics, Zhejiang University, Hangzhou 310027, China*

[6]*HPSynC, Geophysical Laboratory, Carnegie Institution of Washington, 9700 South Cass Avenue, Argonne, Illinois 60439, USA*

[7]*Shanghai Synchrotron Radiation Facilities, Shanghai Institute of Applied Physics, Chinese Academy of Sciences, Shanghai 201204, China*

[8]*These authors contributed equally to this work.*

**Pressure plays an essential role in the induction[1] and control[2,3] of superconductivity in iron-based superconductors. Substitution of a large cation by a smaller rare-earth ion to simulate the pressure effects has raised the superconducting transition temperature $T_C$ to a record high of 55 K in these materials[4,5]. Analogous to the**

**bell-shaped curve of $T_C$ dependence on chemical doping, pressure-tuned $T_C$ typically drops monotonically after passing the optimal pressure[1-3]. Here we report the observations of an unexpected phenomenon in superconducting iron chalcogenides that after the $T_C$ dropped from the maximum of 32 K at 1 GPa and vanished (< 4 K) above 9.5 GPa, a second superconducting region with considerably higher $T_C$ than the first maximum suddenly reemerged above 11.5 GPa. The $T_C$ of the reemerging superconducting phase reaches 48.0-48.7 K for $Tl_{0.6}Rb_{0.4}Fe_{1.67}Se_2$, $K_{0.8}Fe_{1.7}Se_2$ and $K_{0.8}Fe_{1.78}Se_2$, setting a new $T_C$ record for iron chalcogenide superconductors.**

The recent discoveries of superconductivity at 30-32 K in a new family of iron-based chalcogenide superconductors[6-9] $A_{1-x}Fe_{2-y}Se_2$ (A=K, Rb or Cs, with possible Tl substitution) bring new excitement to the field of superconductivity[10]. Some of the most striking features of such superconductors include the unusually large magnetic moments up to 3.3 $\mu_B$ per Fe atom and the Fe-vacancy ordering in the Fe square lattice[11]. How superconductivity with such a high $T_C$ can exist on such a strong magnetic background remains perplexing[10]. It has been established that the superconductivity in strongly correlated electronic systems can be dictated by their crystallographic structure, electronic charge, as well as orbital and spin degree of freedom, which in turn, can be manipulated by controlling parameters including pressure, magnetic field, and chemical composition[12-15]. Pressure is a 'clean' way to tune basic electronic and structural properties without changing the chemistry. High-pressure studies are thus very useful in elucidating mechanisms of superconductivity as well as in searching for new high-$T_C$ superconducting materials.

We studied single crystals of $Tl_{0.6}Rb_{0.4}Fe_{1.67}Se_2$, $K_{0.8}Fe_{1.7}Se_2$ and $K_{0.8}Fe_{1.78}Se_2$ grown by the Bridgman method[6,16,17]. We have conducted both high-pressure resistance and susceptibility measurements to detect superconductivity *in-situ* at high pressures and low temperatures. Figure 1 shows temperature dependence of the electrical resistance at different pressures for $Tl_{0.6}Rb_{0.4}Fe_{1.67}Se_2$ single crystals. Here we define $T_C$ as the intersection of the tangent through the inflection point of the resistive transition with a straight-line fit of the normal state just above the transition. As can be seen, $T_C$ starts at the maximum of 33 K at 1.6 GPa, shifts to lower temperatures at increasing pressures, and vanishes near 9 GPa in our experimental temperature range, which is 300-4 K for our high-pressure resistance measurements (Figure 1a). At slightly higher pressures, however, an unexpected superconducting phase reemerges with an onset $T_C$ as high as 48.0 K at 12.4 GPa (Figure 1b). The sample is not superconducting at pressures higher than 13.2 GPa. We repeated the measurements with new samples in three independent experiments, and the results were reproducible.

To confirm the pressure-induced changes of superconductivity in $Tl_{0.6}Rb_{0.4}Fe_{1.67}Se_2$, we also performed magnetic *ac* susceptibility measurements *in-situ* at high pressures (Figure 2). The value of $T_C$ is taken to be the onset of superconductivity defined by the intersection of a line drawn through the steep slope of the curve and the region of zero slope above the transition. The magnetic study showed that $T_C$ decreased with increasing pressure and vanished at 9.8 GPa in the first superconducting phase SC-I (Figure 2a).

With further increasing pressure, the material enters a new superconducting phase SC-II and its transition temperature reaches 40.2 K at 12.2 GPa (Figure 2b). The magnetic measurements yield $T_C$ values consistent with the resistivity data within the experimental uncertainties. These results provide convincing evidence for the existence of two distinct superconducting phases in $Tl_{0.6}Rb_{0.4}Fe_{1.67}Se_2$.

To investigate whether the pressure-induced reemergence of superconductivity was unique to $Tl_{0.6}Rb_{0.4}Fe_{1.67}Se_2$ or more general among iron chalcogenides, we conducted parallel electrical resistance measurements on $K_{0.8}Fe_{1.7}Se_2$ single crystals, and observed nearly identical behaviour (Figure 3). The initial $T_C$ of 32 K at 0.8-1.6 GPa decreased monotonically with increasing pressure and became undetectable at 9.2 GPa. At a slightly increased pressure, the second superconducting phase of $K_{0.8}Fe_{1.7}Se_2$ reemerged and reached the maximum $T_C$ of 48.7 K at 12.5 GPa. We repeated the experiment six times using six single crystals cut from different batches, and the results were reproducible. We further repeated the measurements with a slightly different composition, $K_{0.8}Fe_{1.78}Se_2$, and again, observed similar pressure-induced behaviour.

We summarized the pressure dependence of $T_C$ of $Tl_{0.6}Rb_{0.4}Fe_{1.67}Se_2$, $K_{0.8}Fe_{1.7}Se_2$, and $K_{0.8}Fe_{1.78}Se_2$ in Figure 4 and Table S1-S4 of Supplementary Information. The diagram clearly reveals two distinct superconducting regions: the initial superconducting phase SC-I and the pressure-induced superconducting phase SC-II. In the SC-I region, $T_C$ is suppressed with applied pressure and approaches zero between 9.2 and 9.8 GPa. At higher pressures, the SC-II region appears, in which the $T_C$ is even higher than the maximum $T_C$ of the SC-I region. The SC-II region has a maximum $T_C$ of 48.7 K for $K_{0.8}Fe_{1.7}Se_2$ and 48.0 K for $Tl_{0.6}Rb_{0.4}Fe_{1.67}Se_2$, setting the new $T_C$ record in chalcogenide superconductors. The SC-II region appears in a narrow pressure range. Unlike the usual parabolic pressure-tuning curve of $T_C$, the high $T_C$ in SC-II appears abruptly above 9.8 GPa and disappears equally abruptly above 13.2 GPa. Intermediate $T_C$<38 K is not observed even with small pressure increment steps of 0.1 GPa. A similar reemergence of superconductivity has been observed in some other strongly correlated electronic systems, such as heavy-fermion[12] and organic systems[20].

Our preliminary high-pressure polycrystalline X-ray diffraction results of the two iron chalcogenides $K_{0.8}Fe_{1.7}Se_2$ and $K_{0.8}Fe_{1.78}Se_2$ confirm that, to the first degree, the basic tetragonal crystal structure persists throughout the pressure range studied (Supplementary Information). Therefore, the disappearance of $T_C$ in SC-I, the reemergence of higher $T_C$ in SC-II, and the final non-superconducting region reflect detailed structural variances within the basic tetragonal unit cell, that are awaiting in-depth investigations with advanced diagnostic probes in the future. For instance, the possible change in magnetic-ordering structures would require high-pressure neutron diffraction, and the possible superlattice and Fe vacancy ordering would require high-pressure single-crystal X-ray structural investigations.

The pressure dependence of $T_C$ in the SC-I region is expected but its mechanism is still heavily debated. Quantum criticalities are thought to play important roles in superconductivity for strongly correlated electronic systems[21]. A characteristic feature of the new iron chalcogenide superconductors is the existence of Fe-vacancies in the Fe-square lattice, ordered by a $\sqrt{5}\times\sqrt{5}$ superstructure[11]. It remains unclear whether pressure could destroy the vacancy ordering at a critical value and drive the materials into a disordered lattice. Detailed structural studies of these superconducting behaviours in the iron chalcogenide superconductors are currently being conducted. The information on magnetic properties at high pressures is highly desired in order to understand the interplay of magnetism and superconductivity in these iron chalcogenides.

The surprising observation of the SC-II region with the reemerging higher $T_C$ is totally unexpected. It will certainly stimulate a great deal of future experimental and theoretical studies to clarify whether the observed reemergence of superconductivity in iron cholcogenides is associated with the quantum critical transition, magnetism, superstructure, vacancy ordering, and spin fluctuation, *etc*.

**Acknowledgments** We thank I. I. Mazin, W. Bao, and T. Xiang for stimulating discussions and J. S. Schilling for the help with *ac* susceptibility technique. This work in China was supported by the NSCF, 973 projects, and Chinese Academy of Sciences. This work in U.S. was supported as part of the EFree, an Energy Frontier Research Center funded by the U.S. Department of Energy, Office of Science, Office of Basic Energy Sciences. HPCAT is supported by CIW, CDAC, UNLV and LLNL through funding from DOE-NNSA, DOE-BES and NSF. APS is supported by DOE-BES.

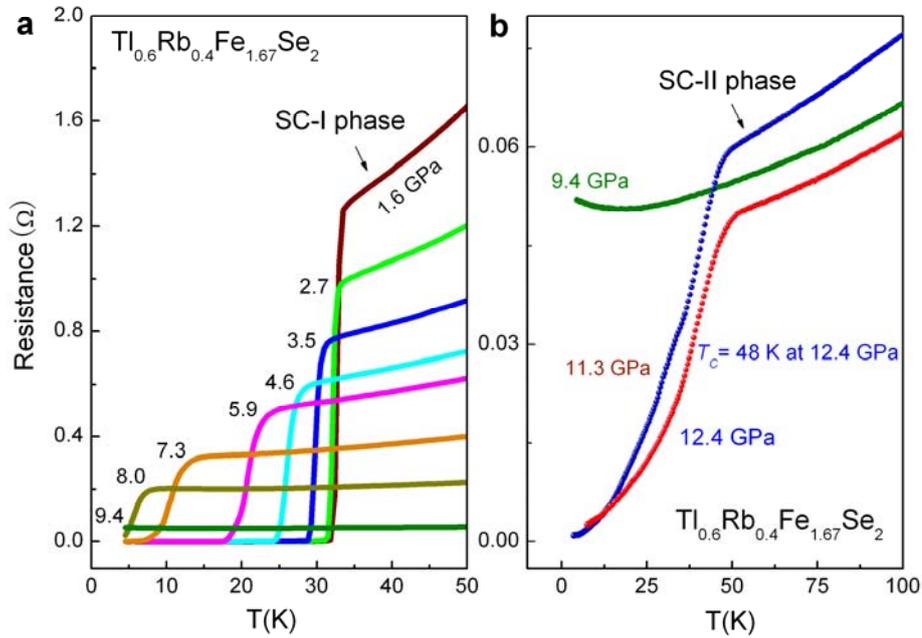

**Figure 1 Temperature-dependence of electrical resistance for $Tl_{0.6}Rb_{0.4}Fe_{1.67}Se_2$ at different pressures. a**, Resistance-temperature curves in the SC-I region up to 9.4 GPa. Superconducting $T_C$ was observed to shift to lower temperature with increasing pressure. Superconductivity disappears at 9.4 GPa. **b**, Electrical resistance curves for the same single crystal at higher pressures. A new superconducting state reemerges upon further compression. The second superconducting phase (SC-II) has a $T_C$ of 48 K which is much higher than the maximum in the initial superconducting phase (SC-I). Cryogenic resistance measurements in terms of standard four-probe method were performed in a diamond-anvil cell made of Be-Cu alloy . Diamond anvils of 600 and 300 μm culet flats were used with 300 μm and 100 μm diameter sample chambers, respectively. Four electrical leads were attached to the single-crystal sample insulated from the rhenium gasket, and loaded into the sample chamber. NaCl powders were employed as a pressure medium. The ruby fluorescence method was used to gauge pressure[21].

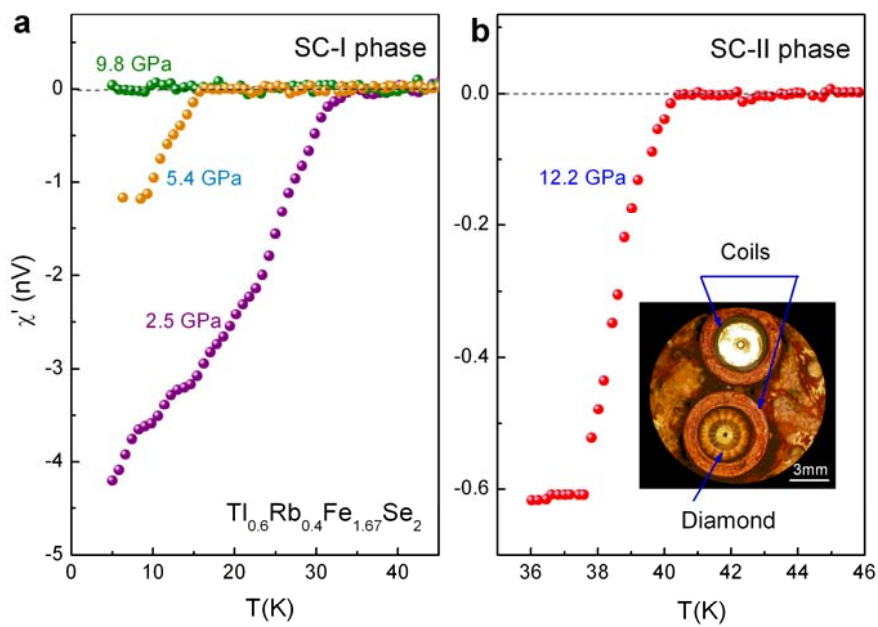

**Figure 2 Temperature dependence of the ac susceptibility for $Tl_{0.6}Rb_{0.4}Fe_{1.67}Se_2$ at different pressures**. **a**, Superconducting transitions observed in the real susceptibility component of the sample at pressure of 2.5, 5.4, and 9.8 GPa in the initial superconducting phase regime (SC-I). The superconducting transition shifts downward to lower temperature with increasing pressure. At 9.8 GPa the susceptibility component remains constant upon cooling down to 4 K, indicating that the sample is no longer superconducting. **b**, The real component of the susceptibility *vs*. temperature for the crystal in the pressure-induced superconducting phase (SC-II) at pressure of 12.2 GPa. Inset: The setup for *ac* susceptibility measurements in a diamond-anvil cell shows a signal coil around the diamond anvils and a compensating coil. The *ac* susceptibilities were detected within a lock-in amplifier[22]. The crystals were loaded into the sample chamber produced from nonmagnetic gaskets with daphne 7373 as pressure medium.

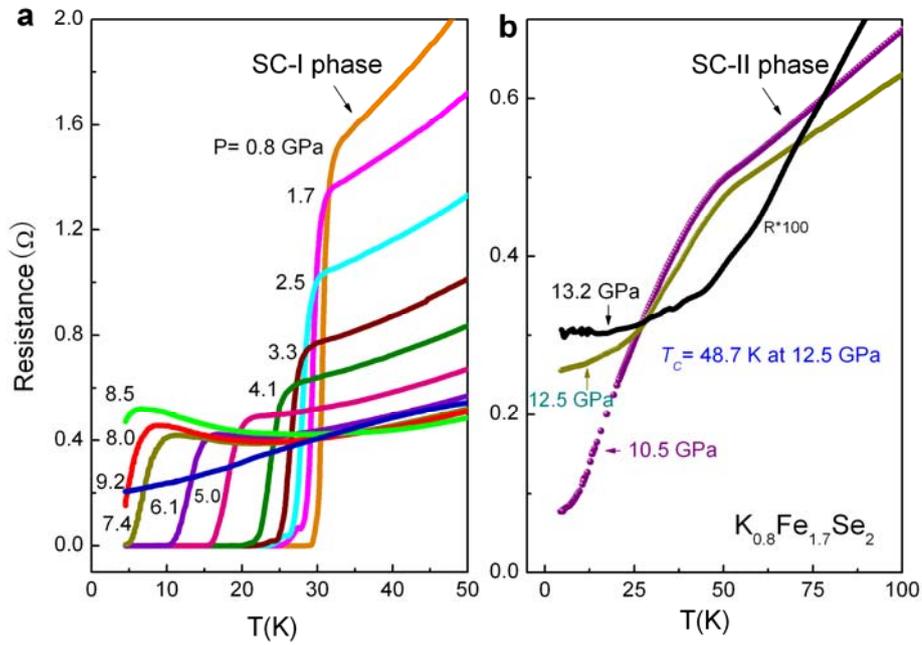

Figure 3 **Temperature dependence of the resistance for $K_{0.8}Fe_{1.7}Se_2$ at different pressures. a**, SC-I. The resistance-temperature curves showing the reduction of the superconducting transition temperature with increasing pressure and its disappearance at 9.2 GPa. **b**, SC-II. The resistance measurements reveal another superconducting phase above 10.5 GPa. The $T_C$ reaches 48.7 K at 12.5 GPa and disappears at 13.2 GPa.

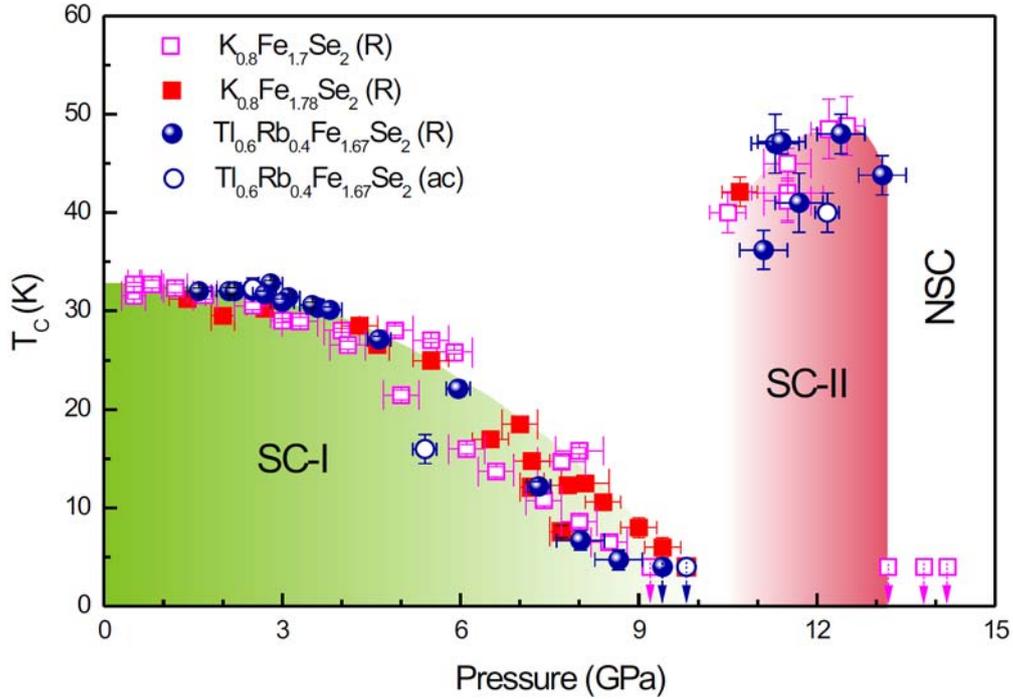

Figure 4 **Pressure dependence of the $T_C$ for $Tl_{0.6}Rb_{0.4}Fe_{1.67}Se_2$, $K_{0.8}Fe_{1.7}Se_2$ and $K_{0.8}Fe_{1.78}Se_2$.** The symbols represent the $P$-$T$ conditions that $T_C$ were observed; symbols with downward arrows represent the absence of superconductivity to the lowest temperature (4 K). All $Tl_{0.6}Rb_{0.4}Fe_{1.67}Se_2$, $K_{0.8}Fe_{1.7}Se_2$ and $K_{0.8}Fe_{1.78}Se_2$ samples show two superconducting regions (SC-I and SC-II) separated by a critical pressure at around 10 GPa. NSC refers to the non-superconducting region above 13.2 GPa. The maximum $T_C$ is found to be 48.7 K in $K_{0.8}Fe_{1.7}Se_2$ at pressure of 12.5 GPa. At higher pressures above 13.2 GPa, the samples are non-superconducting. Error bars, 1 s.d.

## Supplementary information

**Structural evolution with pressure in $K_{0.8}Fe_{1.7}Se_2$ and $K_{0.8}Fe_{1.78}Se_2$**

Figures S1 show the x-ray diffraction patterns for $K_{0.8}Fe_{1.7}Se_2$, and Figures S2, for $K_{0.8}Fe_{1.78}Se_2$ at various pressures at room temperature. The pristine samples were carefully grounded to fine powders for *in-situ* high-pressure diffraction experiments. Angle-dispersive x-ray powder diffraction experiments for $K_{0.8}Fe_{1.7}Se_2$ were performed with a focused monochromatic x-ray beam with wavelength of 0.4246 Å at beamline 16BM-D of HPCAT at the Advanced Photon Source, Argonne National Laboratory and for $K_{0.8}Fe_{1.78}Se_2$ with a focused monochromatic x-ray beam with wavelength of 0.6888 Å at BL15U at Shanghai Synchrotron Radiation Facilities. The patterns can be well refined to the tetragonal structure for these materials.

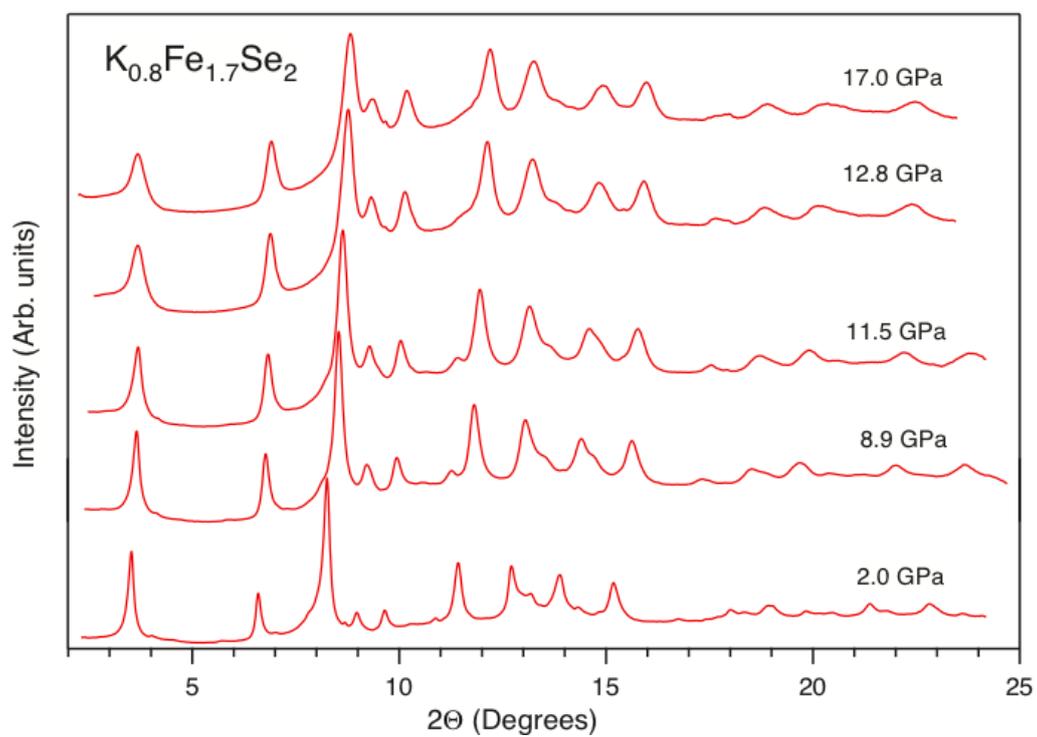

**Figure S1** X-ray powder diffraction patterns of $K_{0.8}Fe_{1.7}Se_2$ at various pressures up to 17.0 GPa.

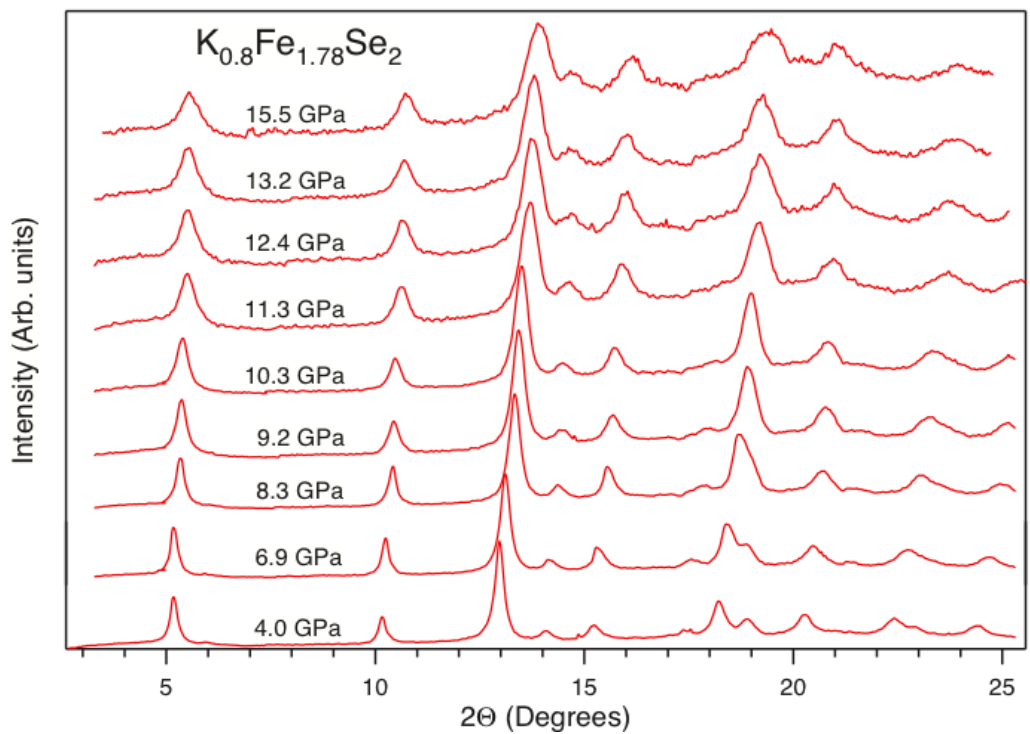

**Figure S2** X-ray powder diffraction patterns of $K_{0.8}Fe_{1.78}Se_2$ at various pressures up to 15.5 GPa.

**Experimental details of the measured $T_C$'s of $Tl_{0.6}Rb_{0.4}Fe_{1.67}Se_2$, $K_{0.8}Fe_{1.7}Se_2$, and $K_{0.8}Fe_{1.78}Se_2$ at high pressures**

The superconductivity of newly discovered iron chalcogenide superconductors is sensitive to pressure. Here we document the experimental details of the measured $T_C$'s and the phases of three compounds at high pressures. In Tables S1-S4, Run # denotes different sample and loading of the same composition. $P_{inc}$ and $P_{dec}$ denote the increasing and decreasing runs, respectively. NO denotes superconductivity is not observed with in our temperature range (300-4K). SC-I and SC-II represent the first and the second superconducting regions and NSC denotes a non superconducting phase.

Table S1 Summary of the measured $T_C$'s and the phases of $Tl_{0.6}Rb_{0.4}Fe_{1.67}Se_2$ from high-pressure resistance measurements.

| Run # | P (GPa) | $P_{inc}$ or $P_{dec}$ | $T_C$ (K) | Phase |
|---|---|---|---|---|
| 1 | 1.60 | $P_{inc}$ | 32.0 | SC-I |
| 1 | 2.20 | $P_{inc}$ | 32.0 | SC-I |
| 1 | 2.50 | $P_{inc}$ | 32.0 | SC-I |
| 1 | 2.81 | $P_{inc}$ | 31.8 | SC-I |
| 1 | 3.10 | $P_{inc}$ | 31.4 | SC-I |
| 1 | 3.50 | $P_{inc}$ | 30.6 | SC-I |
| 2 | 0.88 | $P_{inc}$ | 32.6 | SC-I |
| 2 | 1.65 | $P_{inc}$ | 33.1 | SC-I |
| 2 | 2.72 | $P_{inc}$ | 32.5 | SC-I |
| 2 | 3.60 | $P_{inc}$ | 30.3 | SC-I |
| 2 | 4.63 | $P_{inc}$ | 27.1 | SC-I |
| 2 | 5.96 | $P_{inc}$ | 22.1 | SC-I |
| 2 | 7.32 | $P_{inc}$ | 12.2 | SC-I |
| 2 | 8.02 | $P_{inc}$ | 6.70 | SC-I |
| 2 | 9.40 | $P_{inc}$ | NO | NSC |
| 2 | 11.3 | $P_{inc}$ | 47.0 | SC-II |
| 2 | 11.7 | $P_{inc}$ | 41.0 | SC-II |
| 2 | 12.4 | $P_{inc}$ | 48.0 | SC-II |
| 2 | 0.50 | $P_{dec}$ | NO | NSC |
| 3 | 2.70 | $P_{inc}$ | 31.7 | SC-I |
| 3 | 3.01 | $P_{inc}$ | 30.9 | SC-I |
| 3 | 3.80 | $P_{inc}$ | 30.1 | SC-I |
| 3 | 11.4 | $P_{inc}$ | 47.2 | SC-II |

Table S2 Summary of the measured $T_C$'s and the phases of $Tl_{0.6}Rb_{0.4}Fe_{1.67}Se_2$ from high-pressure *ac* susceptibility measurements.

| Run # | P (GPa) | $P_{inc}$ or $P_{dec}$ | $T_C$ (K) | Phase |
|---|---|---|---|---|
| 1 | 2.50 | $P_{inc}$ | 32.3 | SC-I |
| 1 | 5.40 | $P_{inc}$ | 16.0 | SC-I |
| 1 | 9.80 | $P_{inc}$ | NO | NSC |
| 1 | 12.2 | $P_{inc}$ | 40.2 | SC-II |

Table S3 Summary of the measured $T_C$'s and the phases of $K_{0.8}Fe_{1.7}Se_2$ from high-pressure resistance measurements.

| Run # | P (GPa) | $P_{inc}$ or $P_{dec}$ | $T_C$ (K) | Phase |
|---|---|---|---|---|
| 1 | 4.00 | $P_{inc}$ | 28.0 | SC-I |
| 1 | 5.90 | $P_{inc}$ | 25.8 | SC-I |
| 1 | 8.00 | $P_{inc}$ | 15.8 | SC-I |
| 1 | 5.50 | $P_{dec}$ | 27.6 | SC-I |
| 1 | 11.5 | $P_{inc}$ | 41.2 | SC-II |
| 1 | 14.2 | $P_{inc}$ | NO | NSC |
| 2 | 0.50 | $P_{inc}$ | 32.7 | SC-I |
| 2 | 0.80 | $P_{inc}$ | 32.7 | SC-I |
| 2 | 1.20 | $P_{inc}$ | 32.3 | SC-I |
| 2 | 1.72 | $P_{inc}$ | 31.6 | SC-I |
| 2 | 2.50 | $P_{inc}$ | 30.5 | SC-I |
| 2 | 3.30 | $P_{inc}$ | 28.9 | SC-I |
| 2 | 4.10 | $P_{inc}$ | 26.5 | SC-I |
| 2 | 5.00 | $P_{inc}$ | 21.5 | SC-I |
| 2 | 6.10 | $P_{inc}$ | 16.0 | SC-I |
| 2 | 6.60 | $P_{inc}$ | 13.7 | SC-I |
| 2 | 7.40 | $P_{inc}$ | 10.8 | SC-I |
| 2 | 8.00 | $P_{inc}$ | 8.60 | SC-I |
| 2 | 8.50 | $P_{inc}$ | 6.50 | SC-I |
| 2 | 9.20 | $P_{inc}$ | NO | NSC |
| 2 | 10.5 | $P_{inc}$ | 40.0 | SC-II |
| 2 | 11.5 | $P_{inc}$ | 45.0 | SC-II |
| 3 | 0.50 | $P_{inc}$ | 31.5 | SC-I |
| 3 | 4.90 | $P_{inc}$ | 28.0 | SC-I |
| 3 | 7.70 | $P_{inc}$ | 14.7 | SC-I |
| 3 | 11.5 | $P_{inc}$ | 42.0 | SC-II |
| 3 | 12.5 | $P_{inc}$ | 48.7 | SC-II |
| 3 | 13.2 | $P_{inc}$ | NO | NSC |
| 3 | 13.8 | $P_{inc}$ | NO | NSC |

Table S4 Summary of the measured $T_C$'s and the phases of $K_{0.8}Fe_{1.78}Se_2$ from high-pressure resistance measurements.

| Run # | P (GPa) | $P_{inc}$ or $P_{dec}$ | $T_C$ (K) | Phase |
|---|---|---|---|---|
| 1 | 2.00 | $P_{inc}$ | 29.5 | SC-I |
| 1 | 4.60 | $P_{inc}$ | 26.5 | SC-I |
| 1 | 7.00 | $P_{inc}$ | 18.5 | SC-I |
| 1 | 8.10 | $P_{inc}$ | 12.5 | SC-I |
| 2 | 7.20 | $P_{inc}$ | 12.1 | SC-I |
| 2 | 7.70 | $P_{inc}$ | 7.60 | SC-I |
| 2 | 9.80 | $P_{inc}$ | NO | NSC |
| 3 | 1.40 | $P_{inc}$ | 31.2 | SC-I |
| 3 | 2.70 | $P_{inc}$ | 30.2 | SC-I |
| 3 | 4.30 | $P_{inc}$ | 28.5 | SC-I |
| 3 | 5.50 | $P_{inc}$ | 25.0 | SC-I |
| 3 | 6.50 | $P_{inc}$ | 17.0 | SC-I |
| 3 | 7.20 | $P_{inc}$ | 14.8 | SC-I |
| 3 | 7.80 | $P_{inc}$ | 12.3 | SC-I |
| 3 | 8.40 | $P_{inc}$ | 10.6 | SC-I |
| 3 | 9.00 | $P_{inc}$ | 8.00 | SC-I |
| 3 | 9.40 | $P_{inc}$ | 6.00 | SC-I |
| 3 | 10.7 | $P_{inc}$ | 42.1 | SC-II |
| 3 | 0.50 | $P_{dec}$ | NO | NSC |